\documentclass[aps,prl,amsfonts,nofootinbib,supercriptaddress]{revtex4}

\usepackage{amssymb,amsmath}
\usepackage{graphicx}
\usepackage{epsfig}
\usepackage{bbm}

\usepackage[T1]{fontenc}
\usepackage[latin9]{inputenc}
\usepackage{amssymb}
\usepackage{esint}

\newcommand{\beq}{\begin{equation}}
\newcommand{\eeq}{\end{equation}}
\newcommand{\bea}{\begin{eqnarray}}
\newcommand{\eea}{\end{eqnarray}}
\newcommand{\vc}[1]{{\textbf{#1}}}

\newcommand{\mc}[1]{\mathcal{#1}}

\def\l{\left}
\def\r{\right}

\begin{document}

\title{Parametric Decay of the Curvaton}

\author{K.~Enqvist}
\author{S.~Nurmi}
\affiliation{Department of Physics, University of Helsinki, and Helsinki Institute of Physics, P.O.Box 64,
FIN-00014, University of Helsinki,
Finland}
\author{G.~I.~Rigopoulos}
\affiliation{Helsinki Institute of Physics, P.O.Box 64,
FIN-00014, University of Helsinki,
Finland}

\begin{abstract}

\noindent We argue that the curvaton decay takes place most
naturally by way of a broad parametric resonance. The mechanism is
analogous to resonant inflaton decay but does not require any tuning
of the curvaton coupling strength to other scalar fields. For low scale 
inflation and a correspondingly low mass scale for the
curvaton, we speculate on observable consequences including the
possibility of stochastic gravitational waves

\end{abstract}

\maketitle

\section{Introduction}

The curvaton scenario \cite{curvaton, Lyth, Moroi,earlier} provides
an alternative mechanism for producing the primordial density
perturbations. The curvaton is a light field which is subdominant
during inflation but whose energy density will eventually become an
important (if not actually the dominant) energy component in the
universe. Because of its lightness, during inflation the curvaton
acquires fluctuations which at the time of curvaton decay are
transformed into curvature perturbations.

So far, the decay of the curvaton has mostly been treated as a
perturbative process, see e.g.  \cite{curvatondecay,LUW}, where the
curvaton condensate decays into relativistic particles, determined
by a rate $\Gamma$ and occurring roughly when $H\sim \Gamma$. This
happens at the stage when the curvaton field is oscillating about
the minimun of its potential, whence it behaves like pressureless
dust so that its relative energy density grows with respect to
radiation produced by the inflaton decay. The curvature pertubation
induced by the curvaton decay is mostly adiabatic but can give rise
to a large non-gaussianity \cite{LUW,curvatonng} if the curvaton
energy density is too small\footnote{Strictly speaking, this holds
only for a quadratic curvaton potential, see \cite{sami}.}; this
implies a lower limit on the fraction of the energy density in the
curvaton condensate at the decay time. Another generic feature of 
curvaton models with no mixed inflaton and curvaton perturbations 
is the small amplitude tensor perturbations \cite{Lyth, LUW}; this 
is a consequence of the low inflationary scale required to keep 
the inflaton perturbations negligible. \footnote{ See however \cite{vaihkonenetal}
where it is demonstrated  that the amplitude of CMB-scale tensor 
perturbations in the curvaton  model can be directly proportional to the 
amount of the produced non-gaussianity due to second order effects.}

However, since the curvaton is a coherently oscillating condensate,
it is possible to excite other bose fields in a regime of parametric
resonance, in a manner similar to inflationary preheating
\cite{preheating,linde}. As a consequence, the curvaton decay
process could have important cosmological ramifications for e.g. 
non-gaussianities, which are known to be enhanced \cite{enhance,
arttu} in some preheating scenarios, as well as for stochastic
gravitational waves, as will be discussed below. Preheating is a
violent process in which the occupation numbers of fields coupled to
the curvaton grow exponentially. In effect, dense lumps of decay
products are created very rapidly and randomly in the real space,
and as these highly inhomogeneous concentrates of energy density
collide, there arise large-amplitude, high-frequency gravitational
waves, as has been discussed recently \cite{dufeau, gw}. These do
not leave a signal on the CMB but could possibly be observed in
current or future gravitational wave detectors such as LIGO
\cite{ligo}, LISA \cite{lisa} or BBO \cite{bbo}.

In the present paper we study the possibility of preheating in
generic curvaton models and explore the main consequences of such a
scenario (preheating has been discussed in the context of MSSM flat
direction curvatons in \cite{postma}). Indeed, the curvaton has to
couple to other fields if it is to decay; if it couples to other
scalars, then there would be two options. Either the additional
scalars would be light and shifted from their minima during
inflation, in which case they would simply be additional curvatons,
or they could be massive and driven to their minima. In the first
case one still would have to find a way for the curvaton(s) to
decay. However, in the latter case there can arise a parametric
resonance during curvaton oscillations with the ensuing rapid decay
of the curvaton.

 We find that the preheating
process in the curvaton scenario is qualitatively similar to the
preheating after inflation although the resonance dynamics can be
somewhat altered. This is because the curvaton, unlike the inflaton,
could remain subdominant during the entire preheating stage.
Moreover, in the scenario studied here the backreaction of the
excited bose fields always becomes significant at the end of
preheating whereas in the standard inflationary case the preheating
may terminate already at an earlier stage. Since the gaussian part
of curvaton perturbations on superhorizon scales is not affected by
preheating and the final decay of the curvaton will eventually take
place in a perturbative reheating process, the possible preheating
stage does not alter the standard predictions of the curvaton model
\cite{LUW}. However, interestingly enough we find that preheating
may generate stochastic gravitational waves with an amplitude much
larger than the inflationary tensor perturbations in the curvaton
model, and we discuss the feasibility of observing such signatures
of curvaton preheating. The study of the non-gaussianities is left
for a future work.

\section{The inflationary period}

Preheating, or resonant decay of the curvaton,  may occur in a large
class of curvaton models with several interacting scalar fields, as
we will now discuss. To illustrate the main features of the process,
let us consider a very simple model with the potential
\beq
\label{potentials}
V=V(\phi)+V(\sigma,\chi)~,
\eeq
where $\phi$ is the inflaton, $\sigma$ is the curvaton and $\chi$ is
an additional scalar field which the curvaton decays into.
Note that $\chi$ should not be yet another curvaton.
We assume that the inflationary dynamics
is completely determined by the inflaton field, $H_{*}=H_{*}(\phi)$,
which requires that the energy densities of the curvaton and the
$\chi$ field are much smaller than $H_{*}^2M_{\rm P}^2$. We also assume
that all fields have canonical kinetic terms.

Let us further assume that
  \beq
  \label{pot}
  V(\sigma,\chi)=
  \frac{1}{2}m_{\sigma}^2\sigma^2+\frac{1}{2}g^2\sigma^2\chi^2+\frac{1}{4!}\lambda\chi^4\
  ,
  \eeq
where $g\lesssim 1$ and $\lambda \lesssim 1$ are some coupling
constants. For certain field and coupling values, both $\sigma$ and
$\chi$ could act as the curvaton but this is not the case we wish to
study here. We therefore concentrate on the regime where the
effective mass of the $\chi$ field is larger than the Hubble
parameter $g\sigma \gtrsim H_{*}$ and the contribution to the
curvaton mass coming from the $\chi$ field can be neglected $\chi\ll
m_{\sigma}/g$. In this case the field $\chi$ will not acquire
perturbations during inflation and it simply represents a generic
bose field that is coupled to curvaton but will not itself act as a
curvaton. Under these assumptions, the value of the $\chi$ field is
driven towards zero exponentially fast during inflation.

The curvaton is effectively massless during inflation $m_{\sigma}\ll
H_{*}$ and acquires perturbations proportional to the Hubble
parameter $\delta\sigma \sim H_{*}$. In the standard scenario where
the curvaton is assumed to decay perturbatively at some stage after
inflation, the amplitude of primordial curvature perturbations can
be estimated by $\mc{P}_{\zeta}^{1/2}\sim r H_{*}/\sigma$
\cite{Lyth,LUW}, where $r\sim\rho_{\sigma}/\rho$ measures the
fraction of the curvaton of the total energy density at the decay
time. Since the curvaton is very light, its equation of motion is
approximatively given by
$3H_{*}\dot{\sigma}\simeq-m_{\sigma}^2\sigma$ (for a discussion of
curvaton dynamics in various potentials, see \cite{Dimopoulos}). If
inflation lasts long enough, the curvaton will eventually roll down
to the region in its potential where the quantum effects overcome
the classical drift, $|\dot{\sigma}|\lesssim H_{*}^2$ . The typical
curvaton value during inflation in this case is thus given by
  \beq
  \label{sigma*}
  \sigma_{*}\sim
  H_{*}\l(\frac{H_{*}}{m_{\sigma}}\r)^2\ .
  \eeq
In the limit $r\ll1$, this yields an estimate
  \beq
  \label{r_osc}
  r_{\rm osc} \sim \left(\frac{m_{\sigma}}{M_{\rm
  P}}\right)^2\left(\frac{H_{*}}{m_{\sigma}}\right)^6
  \eeq
for the curvaton contribution to the total energy density at the
beginning of oscillations.

\section{The parametric resonance}

Preheating in the curvaton model can take place when the curvaton
starts to oscillate after inflation and gives a time dependent
effective mass for the $\chi$ field that initially has a vanishing
vev. The situation is analogous to the standard case of preheating
in the conventional inflationary models \cite{preheating,linde}. As
usual, an efficient preheating requires broad resonance bands which
implies the condition
  \beq
  \label{q}
  q\equiv\frac{g^2\sigma^2}{4m_{\sigma}^2} \gg 1
  \eeq
at the beginning of oscillations. In our case, this necessarily
holds as a consequence of the mass condition for the $\chi$ field,
$g\gtrsim H_{*}/\sigma$, combined with the lightness of the curvaton
$m_{\sigma}\ll H_{*}$.

Depending on the fraction of the energy density in the curvaton
field at the time of oscillations, denoted as $r_{\rm osc}$, there
are two different possibilities for realizing the preheating stage.
If the curvaton dominates the energy density at the beginning of
oscillations with $r_{\rm osc}\sim 1$, preheating proceeds exactly
in the same manner as in the standard inflationary case. In the
opposite limit, $r_{\rm osc}\ll 1$, the universe is not dominated by
the oscillating curvaton but by the inflaton decay products assumed
to be homogeneous radiation. This however leads only to minor
differences as compared to the standard case \cite{linde} as we will
discuss below.

In the limit $r_{\rm osc} \sim 1$ the universe is effectively
matter dominated during the curvaton oscillations. The time
evolution of the curvaton is given by
   \beq
  \label{curvaton}
  \sigma(t)=\frac{\sigma_{*}}{m_{\sigma}t}\,{\rm sin}(m_{\sigma}t) \equiv \bar{\sigma}\, {\rm
  sin}(m_{\sigma}t)\ ,
  \eeq
where $\bar{\sigma}\propto t^{-1}$ measures the amplitude of
oscillations. For definiteness, we choose $t_0=\pi/(2m_{\sigma})$ as
the beginning of oscillations and set $a_0=1$. The initial amplitude
is then given by $\bar{\sigma}=2\sigma_{*}/\pi$, where $\sigma_{*}$
is the value during inflation. Using a rescaled field variable
$X_k=a^{3/2}\chi_k$, the equation of motion for the $\chi$ field in
Fourier space reads
  \beq
  \label{X_k}
  \ddot{X}_k+\left(\frac{k^2}{a^2}+g^2\sigma^2\right)X_k=0\ ,
  \eeq
where terms proportional to $H^2 \ll g^2\sigma^2$ and the small
contribution from the quartic term in (\ref{pot}) have been
neglected. The solutions can be formally written as
  \beq
  \label{X_ksol}
  X_k(t)=\frac{\alpha_k(t)}{\sqrt{2\omega}}e^{-i\int^t\omega dt}+\frac{\beta_k(t)}{\sqrt{2\omega}}e^{i\int^t\omega
  dt}\ ,
  \eeq
and the occupation number of a given mode is determined by
$n_k=|\beta_k|^2$. In the broad resonance regime, the occupation
numbers $n_k$ remain constant for most of the time since the mass of
the $\chi$ field is slowly varying and much larger than the curvaton
mass $m_{\sigma}$ \cite{linde}. However, as the oscillating curvaton
crosses the origin during each cycle of oscillation, the $\chi$
field becomes effectively massless and its frequency
$\omega_k=\sqrt{(k/a)^2+g^2\sigma^2}$ changes in a non-adiabatic
fashion $\dot{\omega}_k\gtrsim\omega_k^2$. This can lead to copious
production of $\chi$ particles via a resonant decay of the curvaton
condensate.

The particle production takes place at very short time intervals around times $t_n$ defined by
$\sigma(t_n)=0$. The curvaton value in the vicinity of these
intervals can be approximated by $|\sigma| \simeq \bar{\sigma}_n
m_{\sigma}(t-t_n)$. By substituting this into (\ref{X_k}), the
equation of motion becomes
  \beq
  \label{Xeom}
  \frac{d^2 X_k}{d\tau^2}+(\kappa_n^2+\tau_n^2)X_k=0\ ,
  \eeq
where we have introduced the variables $k_{*}\equiv\sqrt{g
m_{\sigma} \bar{\sigma}}$, $\kappa_n={k}/(ak_{*})$,
$\tau_n=k_{*}(t-t_n)$ and all the time arguments not explicitly
written out are understood to be evaluated at $t_n$. The solutions
for this equation are parabolic cylinder functions
$W(-\kappa_n^2/2;\pm\sqrt{2}\tau_n)$. By comparing them in the
subsequent adiabatic regimes $t_{n-1}<t<t_{n}$ and $t_{n}<t<t_{n+1}$
one can estimate the change of occupation numbers $n_k$ around the
times $t_n$ when the system is non-adiabatic. In the limit of large
occupation numbers, $n_k\gg1$, the result is given by \cite{linde}
  \beq
  \label{nexp}
  n_k^{n+1}=n_k^n\; {\rm exp}(2\pi\mu_k^n)\ ,
  \eeq
where $2\pi\mu_k^n={\rm ln}(1+2e^{-\pi\kappa_n^2}-2{\rm
sin}\theta_n\;e^{-\pi\kappa_n^2/2}\sqrt{1+e^{-\pi\kappa_n^2}}\;)$
and $\theta_n$ is a phase factor that is essentially a random
variable for each resonance time $t_n$. This is why preheating in an
expanding space has a stochastic nature.

The system formally describes scattering of waves in a parabolic
potential. Each resonance time $t_n$, when the curvaton hits the
origin, corresponds to a scattering event and changes the occupation
numbers of the $\chi$ modes. The changes can be significant in the
regime $\pi\kappa_n^2\lesssim 1$ which corresponds to the momentum
band
  \beq
  \label{band}
  \frac{k^2}{a^2(t_n)}\lesssim\frac{k_{*}^2(t_n)}{\pi}=\frac{g m_{\sigma}
  \bar{\sigma}_n}{\pi}\ .
  \eeq
On average, the occupation numbers grow exponentially in this band,
although the exponent $\mu_k$ in (\ref{nexp}) can also be negative
for some scattering events $t_n$, and the comoving particle number
density as a function of time can estimated as
  \beq
  \label{n_t}
  n(t)=\frac{1}{(2\pi a)^3}\int {\rm d^3k}\; n_k(t)
  \sim 10^{-3}\frac{k_{*}^3}{a^3\sqrt{ \mu m_{\sigma} t}}e^{2 \mu m_{\sigma} t}\ .
  \eeq
Here $\mu$ is the maximum of the effective growth index $\mu_k^{\rm
eff}$ obtained from equation (\ref{nexp}) by defining $\mu_k^{\rm
eff}t=\int^t{\rm dt}\mu_k(t)$.

As the modes of the $\chi$ field grow exponentially, the term
$g^2\chi^2\sigma^2$ in the curvaton potential becomes comparable to
the mass term $m_{\sigma}^2\sigma^2$ and the backreaction effects
which eventually terminate the resonance need to be taken into
account. The backreaction becomes important \cite{linde} when the
number density of $\chi$ particles (\ref{n_t}) grows to $n(t_{\rm
br}) \sim {m_{\sigma}^2\bar{\sigma}}/{g}$ which yields an equation
  \beq
  \label{t_br}
  t_{\rm br}\sim \ \frac{1}{4 \mu m_{\sigma}}
  {\rm ln}\left(\frac{10^6 \mu m_{\sigma}^4 t_{\rm br}^3}{g^5
  \sigma_{*}}\right)\ ,
  \eeq
from which $t_{\rm br}$ can be solved. The condition for broad
resonance (\ref{q}) becomes violated when $g\bar{\sigma}(t_{\rm
f})\sim m_{\sigma}$, which yields
  \beq
  \label{t_f}
  t_{\rm f}\sim \frac{g\sigma_{*}}{m_{\sigma}^2}\gtrsim
  \frac{H_{*}}{m_{\sigma}^2}\ .
  \eeq
The last inequality follows from the condition $g\gtrsim H/\sigma$
which is required to keep $\chi$ massive during inflation. The
backreaction generically takes place much before the end of broad
resonance since the above equations yield $t_{\rm br} \ll t_{\rm f}$
for $m_{\sigma}/H_{*}\lesssim 10^{-2}$, assuming the curvaton value
during inflation is given by equation (\ref{sigma*}). The
interaction between the curvaton condensate and the $\chi$ particles
breaks the condensate into curvaton particles which eventually come
to dominate over the $\chi$ particles \cite{linde}. As usual,
preheating needs to be followed by a reheating stage which
eventually yields the hot big bang era.  The simplest assumption
\cite{linde} is that preheating just prepares slightly different
initial conditions for the reheating stage during which the curvaton
decays into standard model fields. In this case the standard
predictions of the curvaton scenario for the gaussian part of
density fluctuations remain largely unaltered. It is possible though
that the non-thermal epoch after preheating could significantly
affect the picture and lead to serious constraints on the curvaton
scenario; this question should be addressed numerically by studying
the thermalization process. However, even if the gaussian density
fluctuations would not be affected, curvaton preheating leads to
production of gravitational waves which could be detectable as
discussed below. In addition, although we do not discuss this
possibility further here, we note that preheating could also
generate a significant level of non-gaussian density perturbations
as can happen in the usual inflationary case \cite{enhance, arttu}.

If the curvaton is subdominant at the beginning of oscillations,
$r_{\rm osc}\ll1$, the above discussion gets slightly modified but
the qualitative results remain the same. The exact time evolution of
the curvaton in the radiation dominated background is given by
Bessel functions but the solution is well approximated by
  \beq
  \label{appr}
  \sigma(t)\simeq\frac{\sigma_{*}}{(m_{\sigma} t+\frac{\pi}{8})^{3/4}}\,{\rm sin}\left(m_{\sigma} t+\frac{\pi}{8}\right)\
  .
  \eeq
This differs from the matter dominated case (\ref{curvaton}) only by
a phase and the slightly slower decrease of the amplitude
$\bar{\sigma}\propto t^{-3/4}$. It is therefore rather evident that
the analysis of preheating will not significantly differ for the
matter dominated case. The equation of motion for the $\chi$ field
can be expressed exactly in the same form as equation (\ref{Xeom})
except that the time evolution of the variables $\kappa_n$ and
$\tau_n$ differs from the matter dominated case due to the different
evolution of the scale factor $a(t)$ and the curvaton amplitude
$\bar{\sigma}$. This does not affect the stochastic nature of the
successive scatterings at times $t_n$ and the occupation numbers of
the $\chi$ field grow essentially at the same rate as before. We
therefore expect that the number density of $\chi$ particles in the
$r_{\rm osc}\ll1$ limit can be estimated by the same equation
(\ref{n_t}) as in the case $r_{\rm osc}\sim 1$. This conclusion can
be checked by solving the equations of motion for the curvaton and
the $\chi$ field numerically. As illustrated in Fig. $(1)$, the
results in the matter and radiation dominated cases indeed coincide
quite precisely in the regime where the backreaction of $\chi$
particles can be neglected.
\begin{figure}
    \includegraphics[width=9cm]{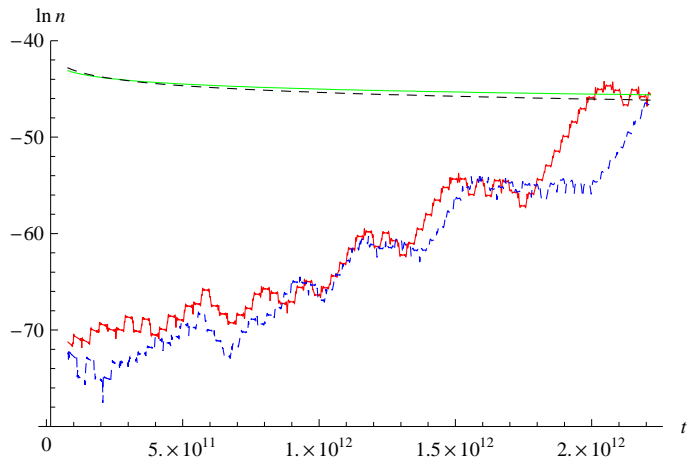}
    \caption{Comparison of resonant decay of the curvaton in radiation and matter dominated backgrounds.
    The figure shows $\ln {n}$ for the $\chi$ field in the cases r=1 (dashed lines)
    as well as $r<1$ (solid lines). It also shows the limits above which
    backreaction becomes important in both cases.
    The parameters for these plots are ($M_P=1$), $m=10^{-10}$, $g=10^{-4}$,
    and $\sigma_*=0.0002$. The onset of  backreaction takes place after about 30 - 35
    oscillations of the curvaton which corresponds to  growth index $\mu \simeq 0.07$.}
\end{figure}
The time when the number density of the $\chi$ particles has grown
large enough to yield a significant backreaction can be estimated as
before using equation (\ref{n_t}) and the condition $n(t_{\rm
br})\sim m_{\sigma}^2\bar{\sigma}/g$. Due to the smaller suppression
by the scale factor in (\ref{n_t}), the backreaction will occur
slightly earlier than in the matter dominated case, as also seen in
Fig. $(1)$,  but the effect is very small. In analyzing the dynamics
after the onset of backreaction, the expansion of the universe can
be neglected \cite{linde} and the radiation domination should not
affect the situation here either.

\section{Gravity waves}

Since the decay of the curvaton can take place in a regime of
parametric resonance, as was demonstrated above, it is natural to
expect that gravitational waves will also be produced when it
occurs. Indeed, the physics is similar as in the inflationary
preheating: an oscillating field excites the production of particles
in other bose fields at the times when their effective frequencies
change non-adiabatically. According to \cite{dufeau}, in such
circumstances the generation of gravitational waves is inevitable.
In what follows we make extensive use of the analysis in
\cite{dufeau} which can be directly applied to our case with
appropriate modifications.

Let us first reiterate the salient features of production of
stochastic gravitational waves, which can be represented as the
transverse-traceless part of the metric perturbations
    \beq
    ds^2=a^2(\eta)\left[-d\eta^2+\left(\delta_{ij}+h_{ij}\right)dx^idx^j\right]\,,
    \eeq
where $\partial_ih_{ij}=h_{ii}=0$ and $d\eta=a^{-1}dt$.
Gravitational waves are sourced by the transverse-traceless part of
the energy momentum tensor $T^{TT}_{ij}$ \footnote{We are
considering here classical gravitational waves generated by the
motion of matter and not vacuum fluctuations amplified by the cosmic
expansion.}
    \beq\label{tensor}
    \partial_\eta^2\bar{h}_{ij}(\vc{k})+\left(k^2-\frac{\partial_\eta^2a}{a}\right)\bar{h}_{ij}(\vc{k})=16\pi G a(\eta) T^{TT}_{ij}(\vc{k})\,,
    \eeq
where $\bar{h}_{ij}=ah_{ij}$. In the case of radiation domination
$\partial_\eta^2a=0$ so equation (\ref{tensor}) can be solved with
the appropriate Green functions of the harmonic oscillator. Since we
will be considering modes with $k\gg aH$ the same applies to a good
approximation for the matter dominated case. The energy density in
gravitational waves for subhorizon modes can then be written as
\cite{dufeau}
    \beq\label{rhoGW}
    \rho_{\rm gw}=\frac{1}{32\pi G a^4}\langle \partial_\eta \bar{h}_{ij}\partial_\eta \bar{h}_{ij}\rangle\,.
    \eeq
Considering the gravitational waves being produced up to a time
$\eta_{\rm f}$ we have
    \beq
    \left(\frac{d\rho_{\rm gw}}{d\ln k}\right)_{\eta>\eta_{\rm f}}=\frac{S_k(\eta_{\rm f})}{a^4(\eta)}\,,
    \eeq
where $S_k$ for a universe filled with a scalar field $\chi$ reads
    \beq
    \label{S}
    S_k(\eta_{\rm f})=\frac{2G}{\pi}k^3 \int \frac{d^3\vc{p}}{\left(2\pi\right)^3}p^4 \sin^4\left(\hat{\vc{k}},\hat{\vc{p}}\right)
    \left\{\left|\int^{\eta_{\rm f}} \!\!\!\!d\eta\, a(\eta)\cos(k\eta)\chi_{k}(\eta)\chi_{|\vc{k}-\vc{p}|}(\eta)\right|^2 +
    \Big|\cos(k\eta)\leftrightarrow\sin(k\eta)\Big|^2\right\}\,.
    \eeq
Here $(\hat{\vc{k}},\hat{\vc{p}})$ is the angle between the unit
vectors and the factors $\sin(k\eta)$ and $\cos(k\eta)$ come from
the harmonic oscillator's Green functions. Taking into account the
expansion of the universe from $\eta_{\rm f}$, the present day
energy density spectrum of the gravitational waves becomes
    \beq
    \label{omega_eq}
    \Omega_{\rm gw} h^2 =  \frac{S_k(\eta_{\rm f})}{a_{\rm osc}^4\rho_{\rm osc}}\left(\frac{g_0}{g_{\rm osc}}\right)^{\frac{1}{3}} \Omega_{\rm r 0}h^2\,,
    \eeq
where we have assumed radiation domination, $r_{\rm osc}\ll 1$,
throughout the generation of gravitational waves for simplicity, $g$
refers to the number of relativistic species and $\Omega_{\rm r
0}h^2$ is the energy density of radiation today.

According to equation (\ref{band}) the typical physical momenta
amplified during preheating are $k_{*}\sim m_{\sigma}q^{1/4}$ and
the gravitational wave signal will also be peaked around $k_{*}$.
 The corresponding frequency today is \cite{dufeau}
    \beq
    \label{f}
    f_k \sim \frac{k_{*}}{\rho_{\rm p}^{1/4}}\,\,10^{10}
    \,{\rm Hz} \sim g^{1/2} r_{\rm osc}^{1/4} a_{\rm p}^{3w/4}\,\,
10^{10}\, {\rm Hz}
    \eeq
where the subscript "p" refers to the time when the peak was formed
and $w$ is the equation of state parameter, $w=1/3$ for $r_{\rm
osc}\ll 1$ and $w=0$ for $r_{\rm osc}\sim 1$. The scale factor at
the beginning of oscillations has been normalized to unity as
before. Using equation (\ref{r_osc}), which holds for models where
the curvaton value during inflation is given by (\ref{sigma*}), and
taking into account the constraint $g\sigma_{*}\gg H_{*}$ required
to keep $\chi$ massive, one finds the bound
  \beq
  \label{f_lower}
  f_k \gtrsim \sqrt{\frac{H_{*}}{M_P}}\,\, 10^{10} \,
  {\rm Hz}\ .
  \eeq
The frequencies accessible to direct detection experiments range
from $f\sim 10^{-4}\,{\rm Hz}$ (LISA) to $f\sim 10^{3}\,{\rm Hz}$
(LIGO). According to equation (\ref{f_lower}), the gravitational
waves produced in the curvaton preheating can fit this observable
window only if the inflationary scale is low enough. If the curvaton
would decay perturbatively and if there are no significant changes
in its effective mass after inflation, the inflationary scale is
constrained by $H_{*} \gtrsim 10^7 {\rm GeV}$ \cite{lythbound}. This
would yield $f_k \gtrsim 10^4 {\rm Hz}$ which unfortunately slightly
escapes the observable window. However, equation (\ref{f_lower}) is
a rather crude estimate and a proper numerical computation is
required to really find out the actual frequency of the
gravitational waves. Moreover, the bound $H_{*} \gtrsim 10^7 {\rm
GeV}$ does not directly apply here since the effective curvaton mass
increases during the non-linear stages of preheating and the highly
non-thermal epoch after the end of preheating may also significantly
affect the expansion history of the universe as compared to the
perturbative curvaton decay.

Having discussed the frequencies of the gravitational waves, we are
now ready to return to the question of their amplitude.
Concentrating first on the linear regime where the backreaction of
$\chi$ particles can be neglected, we may use equation (\ref{S}).
Gravity waves will only be produced when the frequency of the $\chi$
modes varies non-adiabatically, near the zeros of the oscillatory
curvaton. Near those zeros the equation of motion for the $\chi$
field is given by (\ref{Xeom}) (remember $X=a^{3/2}\chi$) and its
solutions are parabolic cylinder functions
$W(-\kappa_n^2/2;\pm\sqrt{2}\tau_n)$
    \beq
    X_k^n=A_k^nW(-\kappa_n^2/2; \sqrt{2}\tau_n) + B_k^nW(-\kappa_n^2/2; -\sqrt{2}\tau_n)\ .
    \eeq
This may be matched to the adiabatic solutions (\ref{X_ksol}) in the
region where the two solutions overlap. Assuming that the occupation
numbers (\ref{nexp}) always grow with the fastest possible rate,
this yields up to a phase factor \cite{dufeau}
    \beq
    X_k^n= \sqrt{\frac{2^{1/2}}{\xi_k}}\frac{\beta_k^n}{\sqrt{k_*}}\,W(-\kappa_n^2/2; -\sqrt{2}\tau_n)\ ,
    \eeq
where $\xi_k\equiv \sqrt{1+e^{-\pi\kappa^2}}-e^{-\pi\kappa^2/2}$.
Using this form and switching to the comoving time in (\ref{S}) we
find
    \beq\label{S_k}
    S_k(\tau_{\rm f})=\frac{m^6 G }{8\pi^3 q\, a^6(\tau_{\rm f})}   K^3 \int\limits^1_{-1}du \left(1-u^2\right)^2
    \int dP P^6 \xi_p\,\xi_{|\vc{k}-\vc{p}|}\,\,n_p\,(\tau_{\rm f})n_{|\vc{k}-\vc{p}|}(\tau_{\rm f})
    \,\,\left(I_{\rm c}^2+I_{\rm s}^2\right)\ ,
    \eeq
where $K=k/m_{\sigma}$ and $I_{\rm s}$, $I_{\rm c}$ are integrals of
the parabolic cylinder functions. Except for the prefactors in front
of the integral and the exact behaviour of the occupation numbers,
this result coincides with equation (78) in \cite{dufeau} derived
for the gravitational waves produced in the linear regime of
preheating after $\lambda \phi^4$ inflation.

Ignoring the differences in $n_k$ for a while, we may then relate
the amplitude of the gravitational waves (\ref{omega_eq}) obtained
in \cite{dufeau}, $\left(\Omega_{\rm gw}h^2\right)_{\lambda\phi^4}$,
to the amplitude in our case as
   \beq
   \label{omega_est}
   \Omega_{\rm gw}h^2 \sim
   10^{27}\left(\frac{m_{\sigma}}{M_P}\right)^4 \left(\Omega_{\rm gw}h^2\right)_{\lambda\phi^4}\ ,
   \eeq
where subleading contributions coming from different values (and
powers) of the scale factor and $q$ have not been included. For a curvaton mass 
at the TeV scale, the amplitude in (\ref{omega_est}) is suppressed at least by a factor
$\sim 10^{-33}$ or so when compared to $\left(\Omega_{\rm
gw}h^2\right)_{\lambda\phi^4}$. From figure 5 of \cite{dufeau} one
finds $\left(\Omega_{\rm gw}h^2\right)_{\lambda\phi^4} \lesssim
10^{-13}$ for the linear regime in $\lambda\phi^4$ preheating. Using
this in (\ref{omega_est}), we see that the amplitude is hopelessly
below the lower limit of observational sensitivity $\Omega_{\rm
gw}h^2\sim 10^{-18}$ (BBO). This conclusion holds even when
differences in the occupation numbers between our case and the
$\lambda\phi^4$ model are taken into account. Assuming that both
models would exhibit similar effective growth rates $\mu$ but the
number of oscillations would be larger by a factor $\Delta N$ in the
curvaton case, the occupations numbers in the curvaton preheating
would be enhanced by a factor ${\rm exp}(4\pi\mu\Delta N)$ as
compared to the $\lambda\phi^4$ case. This would enhance the
amplitude (\ref{omega_est}) by ${\rm exp}(8\pi \mu \Delta N)$ but it
would still be difficult to compensate for the suppression
$(m_{\sigma}/M_P)^4$ in (\ref{omega_est}). For example, for the case
$\mu\simeq 0.07$ shown in Fig. $(1)$ one would need $\Delta N\gtrsim
37$, which is more than the number of oscillations taking place
during the linear stage, to climb above the threshold $\Omega_{\rm
gw}h^2\sim 10^{-18}$. We therefore expect that no directly
observable gravitational waves will be produced in the {\it linear}
regime of the curvaton preheating.

However, the non-linear stages that follow the linear part of the
resonance can give rise to significant production of gravitational
waves and the dominant contribution to the total energy density of
gravitational waves is typically generated during this stage
\cite{gw,dufeau}. Moreover, the non-linear stages are particularly
important in the curvaton model since the backreaction effects
always become important at the final stages of preheating as
discussed above. According to \cite{dufeau}, the gravitational wave
amplitude from these stages as observed today can be estimated as
    \beq\label{gw-nl}
    \Omega_{\rm gw}h^2 \sim 10^{-6} (R_{\rm p}H_{\rm p})^2\,,
    \eeq
where $R_{\rm p}$ is the characteristic physical scale where
parametric resonance is efficient and $H_{\rm p}$ is the expansion
rate at the time when it occurs.  Setting $R_{\rm p} \sim 1/k_*$
and using equations (\ref{sigma*}) and (\ref{r_osc}) we find
    \beq
    \label{omega_nl}
    \Omega_{\rm gw}h^2 \sim \frac{10^{-6}}{g x^2}\left(\frac{m_\sigma}{H_{*}}\right)^3 \lesssim \frac{10^{-6}}{x^2}\frac{m_{\sigma}}{H_{*}}\,
    \eeq
where we have defined $x\equiv m/H_{\rm p}$ and also used the
condition $g\sigma_{*} \gg H_{*}$ in the last step. The parameter
$x$ is proportional to the number of oscillation cycles and
contributes a factor of few tens. The amplitude of the gravitational
waves is therefore typically large enough to be in the detectable
range, provided that the mass of the curvaton is not vastly smaller
than the inflationary scale $H_{*}$. Curvaton models with a
sufficiently low inflationary scale, required to fulfill the
frequency condition (\ref{f_lower}), might therefore produce a
detectable signal of stochastic gravitational waves. This is an
interesting outcome in light of supersymmetric models \cite{flat},
for example, where the natural curvaton mass would be around the
${\rm TeV}$ scale. If the inflationary scale is not too much higher,
these models could yield directly observable gravitational waves.
However, in order to really confirm the qualitative outcome as well
as to find out the exact relation between properties of the
gravitational waves and the curvaton mass and inflationary scale,
one would need to resort to a numerical study.

\section{Discussion}

Up to now, discussions of curvaton dynamics have mostly focused on
perturbative decay of the curvaton condensate. Here we point out
that if the curvaton is about to decay directly into bosonic degrees
of freedom, it is highly likely that there will be a
non-perturbative preheating stage before the perturbative decay and
reheating. The curvaton preheating is qualitatively similar to the
conventional inflationary preheating although minor differences can
arise due to the fact that the curvaton does not need to dominate
the universe during the preheating stage. We have also found that
the end of the curvaton preheating is always dominated by the
effects of backreaction of the amplified light bose fields. This is
different from the inflationary case, where the importance of the
backreaction depends on the parameters. Although we have not
discussed the details of backreaction, we expect to find no
significant differences as compared to the inflationary case here
either. This is because the possible differences between the
curvaton preheating and the inflationary preheating stem from
differences in the expansion of the universe which are unimportant
for the final stages of preheating. However, to complete the
analysis, one should study numerically the thermalization of the
universe after the end of preheating. Since the curvature
perturbations generated in the curvaton scenario are sensitive to
the evolution history before the curvaton decay, these stages might
potentially affect even the gaussian perturbations and therefore
yield interesting constraints on the curvaton model itself.

Numerics is also needed to resolve the issue of stochastic
gravitational waves. Based on simple estimates we have
postulated that models with a sufficiently low inflationary scale
could produce gravitational waves with directly detectable
frequencies and amplitudes. For example, supersymmetric models
\cite{flat} with the curvaton mass around the TeV scale could give
rise to observable gravitational waves provided that the
inflationary scale is not several orders of magnitude higher.
However, our results are still preliminary and a full numerical
simulation, subject to a future research, is required before any
firm conclusions can be drawn.

Likewise, the possible enhancement of the non-gaussianities in
curvaton preheating is best addressed by numerical simulations.
Since curvaton preheating is very similar to the inflaton
preheating, we may expect enhancements similar as has been found
numerically by Chambers and Rajantie \cite{arttu}. It therefore
seems that curvaton decay by a parametric resonance, in addition to
being a natural stage in the curvaton scenario, could also have
interesting observational consequences.

\vskip30pt

\centerline{\textbf{Acknowledgements}}
\vskip10pt
We wish to thank J.~F.~Dufaux for critical comments as well as A.~Rajantie 
and A.~Tranberg for illuminating discussions. This work
was supported by the EU 6th Framework Marie Curie Research and
Training network "UniverseNet" (MRTN-CT-2006-035863) and partly by
Academy of Finland grant 114419. S.N. is supported by the GRASPANP
Graduate School.

\end{document}